 \documentclass[final,5p,times,twocolumn]{elsarticle}

\usepackage{amssymb}
\usepackage{lipsum}
\usepackage[hyphens]{url}  % Better hyphenation in URLs
\usepackage{breakurl}
\usepackage{hyperref}
\usepackage{etoolbox} 
\usepackage{balance}
\usepackage{amsmath}
\usepackage{array}
\usepackage{booktabs}
\usepackage{tikz}
\usetikzlibrary{shapes.geometric, arrows.meta, positioning}
\journal{Pre-print}

\newcommand{\orcidlink}[1]{\textsuperscript{\href{https://orcid.org/#1}{\includegraphics[scale=0.2]{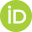}}}}

\begin{document}

\begin{frontmatter}



\title{Choosing the Right Engine in the Virtual Reality Landscape}


\author[]{Santiago Berrezueta-Guzman\,\orcidlink{0000-0001-5559-2056}}
\author[]{Stefan Wagner\,\orcidlink{0000-0002-5256-8429}}
\affiliation[]{organization={Technical University of Munich},
           city={Heilbronn},
            country={Germany}}

\begin{abstract}
Virtual reality (VR) development relies on game engines to provide real-time rendering, physics simulation, and interaction systems. Among the most widely used game engines, Unreal Engine and Unity dominate the industry, offering distinct advantages in graphics rendering, performance optimization, usability, resource requirements, and scalability. This study presents a comprehensive comparative analysis of both engines, evaluating their capabilities and trade-offs through empirical assessments and real-world case studies of large-scale VR projects. The findings highlight key factors such as rendering fidelity, computational efficiency, cross-platform compatibility, and development workflows. These provide practical insights for selecting the most suitable engine based on project-specific needs.
Furthermore, emerging trends in artificial intelligence (AI)-driven enhancements, including Deep Learning Super Sampling (DLSS) and large language models (LLMs), are explored to assess their impact on VR development workflows. By aligning engine capabilities with technical and creative requirements, developers can overcome performance bottlenecks, enhance immersion, and streamline optimization techniques. 

This study serves as a valuable resource for VR developers, researchers, and industry professionals, offering data-driven recommendations to navigate the evolving landscape of VR technology.

\end{abstract}




\end{frontmatter}

\section{Introduction}

Virtual Reality (VR) has revolutionized human-computer interaction, enabling immersive experiences across various domains such as medicine, education, training, and entertainment \cite{zhang2024innovative, checa2020review}. At the core of VR development are game engines, which provide the computational framework for real-time rendering, physics simulation, user interaction, and cross-platform deployment. Initially designed for video games, these engines have evolved into versatile development platforms that support complex simulations, industrial applications, and high-fidelity virtual environments \cite{kleinschmidt2016evaluation}.

Among the leading game engines for VR development, Unreal Engine, and Unity dominate the landscape due to their robust feature sets, extensive ecosystems, and widespread industry adoption.

Unreal Engine, developed by Epic Games, is recognized for its high-fidelity graphics, advanced rendering techniques, and real-time physics simulation. It is widely used in AAA game development, film production, architectural visualization, and industrial simulations. It is particularly suited for projects that require photo-realistic visuals and high computational performance \cite{unreal_nanite, tan2024mastering}. Unreal Engine integrates technologies such as Lumen for real-time global illumination, Nanite for virtualized geometry, and advanced ray tracing capabilities, making it a powerful choice for cutting-edge VR experiences.

Conversely, Unity is known for its flexibility, accessibility, and broad cross-platform support. It provides a lightweight framework that allows developers to create applications for various devices, including PCs, consoles, mobile platforms, and WebGL \cite{unity_multiplatform}. Unity's modular architecture, combined with tools like the Universal Render Pipeline (URP) and High Definition Render Pipeline (HDRP), enables scalable performance optimizations \cite{mercan2017evaluating}. Its user-friendly development environment, large asset store, and strong developer community make it a preferred engine for indie game developers, educational applications, and mobile VR experiences.

Despite widespread adoption, game engines are often misunderstood or undervalued in VR development. However, they serve as the backbone of VR applications, providing the computational foundation for immersive experiences. These engines offer several key functionalities:

\textbf{Graphics Rendering.} Game engines transform complex 3D models and environments into high-quality visual outputs, significantly influencing the user's sense of immersion \cite{anderson2008putting, kumar2020graphics}.

\textbf{Physics Systems.} By simulating collision, gravity, and friction behaviors, game engines provide the realism or creative flexibility necessary for enhancing user engagement or meeting specific application requirements \cite{jungherr2022extended, matej2016virtual}.

\textbf{Performance Optimization.} Maintaining high frame rates is critical for VR applications to prevent motion sickness and deliver smooth user experiences, especially under demanding visual requirements. Game engines are pivotal in achieving this optimization \cite{lavalle2023virtual}.

\textbf{Multi-Platform Support.} Modern game engines facilitate content deployment across various platforms, including PCs, mobile devices, game consoles, and VR headsets. This cross-platform compatibility allows developers to focus on content creation rather than platform-specific complexities \cite{ciekanowska2021comparative}.

\textbf{Development Efficiency.} Integrated development environments, asset stores, scripting tools, and active developer communities streamline the VR development process, enabling rapid prototyping and optimization of immersive experiences \cite{mercan2017evaluating}.

This paper presents a comprehensive comparative analysis of Unreal Engine and Unity in the context of VR development, focusing on their graphics capabilities, performance optimization techniques, usability, hardware requirements, and scalability. The goal is to provide VR developers, researchers, and industry professionals with practical insights into each engine's strengths, limitations, and trade-offs, enabling informed decision-making based on project requirements.

Unreal Engine and Unity were selected for this study due to their widespread adoption, robust feature sets, and proven versatility across both commercial and academic VR applications. Their dominance in the VR development landscape makes them the most relevant platforms for assessing current capabilities, challenges, and future directions in immersive technology.

While previous works have provided theoretical comparisons between Unreal Engine and Unity, few offer a data-driven and future-oriented analysis grounded in recent engine versions and emerging technologies such as AI-driven rendering and large language models. This study fills that gap by incorporating quantitative benchmarks, real-world case studies, and practical insights aligned with current trends in VR development, offering a holistic resource for developers navigating the evolving VR landscape.

The structure of this paper is organized as follows: Section~\ref{M} details the methodology used to evaluate the selected engines, focusing on critical parameters such as graphics rendering, performance optimization, and usability. Section~\ref{TCA} examines the technical capabilities of Unreal Engine and Unity. Section~\ref{ACS} presents case studies to illustrate practical applications and workflows. Section~\ref{CA} offers a comparative analysis, emphasizing the strengths and limitations of each engine. Section~\ref{F} explores future trends, including AI integration and advancements in rendering. Finally, Section~\ref{C} concludes with key findings and recommendations for selecting the ideal game engine for VR projects.

\section{Methodology}\label{M}

This study employs a structured comparative evaluation framework to assess the suitability of Unreal Engine and Unity for VR development. The evaluation is based on three fundamental domains: graphics rendering, performance optimization, and usability. These domains were selected due to their critical role in determining VR applications' overall quality, immersion, and user experience.

\subsection{Evaluation Criteria}

The evaluation criteria for this study were selected based on an extensive literature review and because of their key role in shaping the overall quality of VR applications and user experience.

\textbf{1. Graphics Rendering:} This criterion assesses the engines' capabilities in producing realistic visuals. Factors considered include lighting techniques, texture quality, shader functionality, and advanced rendering features such as ray tracing and global illumination.

\textbf{2. Performance Optimization:} Performance in VR is crucial to maintain high frame rates and minimize motion sickness. This evaluation examines optimization techniques like Level of Detail (LOD)\footnote{LOD is a rendering technique that dynamically adjusts the complexity of 3D models based on their distance from the camera.} management, texture streaming, and hardware utilization for real-time rendering.

\textbf{3. Usability:} Usability influences development efficiency and the learning curve for new users. This criterion includes the availability of development tools, scripting language support, documentation, asset stores, and overall workflow efficiency.

\begin{figure}[h!]
    \centering
    \includegraphics[height=12cm]{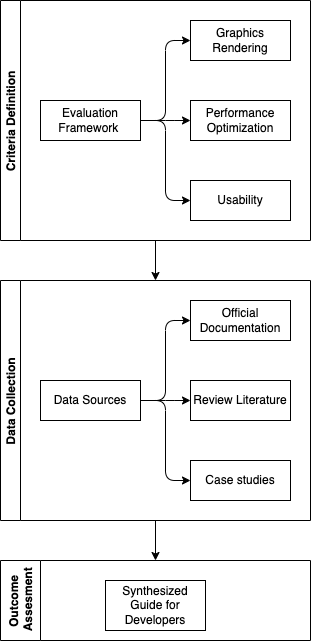}
    \caption{Framework of Research Methodology and Analytical Process}
    \label{fig: scheme}
\end{figure}

Figure \ref{fig: scheme} provides a visual representation of the evaluation process, outlining the relationships between criteria definitions, data sources, and outcome assessment.

\subsection{Literature Review}

The research methodology incorporates a systematic literature review, case study analysis, and data synthesis to assess each game engine's performance. The literature review was conducted primarily using the Scopus database, complemented by official engine documentation. The search strategy included keywords such as "Game Engine and Graphics Rendering," "Performance Optimization for Game Engine," and "Game Engine Usability."

\begin{figure}[h!]
    \centering
    \includegraphics[height=8.2cm]{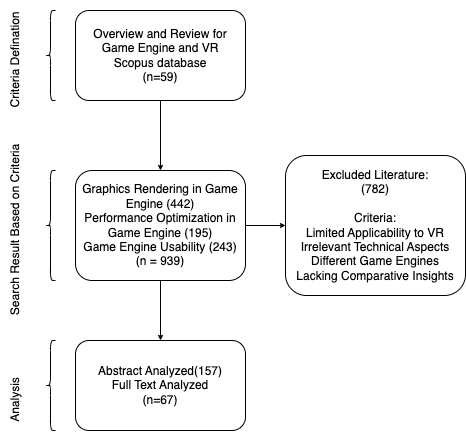}
    \caption{Workflow for Literature Screening and Analytical Process}
    \label{fig: flowchart}
\end{figure}

The initial search yielded 939 publications, which underwent a systematic screening process to ensure relevance. In total, 782 papers were excluded due to their irrelevance to VR applications, focus on non-target game engines and insufficient comparative analysis. After reviewing 157 abstracts, 67 papers were selected for full-text analysis. The literature screening and analytical processing workflow is detailed in Figure \ref{fig: flowchart}.

\subsection{Documentation Analysis}

Following the literature review, a comparative analysis was conducted based on empirical data from case studies of large-scale VR projects. Engine capabilities were evaluated using official documentation and technical benchmarks. The findings were synthesized to highlight the strengths and limitations of each engine, offering practical insights for developers in selecting the most suitable tool for their VR projects.

This structured methodology ensures a rigorous and objective assessment, providing developers with an evidence-based approach to engine selection for VR applications.

\section{Technical Capabilities Analysis}\label{TCA}

This section evaluates the technical capabilities of Unreal Engine and Unity in three fundamental domains crucial to VR development: graphics rendering, performance optimization, and usability. 
These domains collectively determine the visual quality, immersive experience, and ease of development for VR applications.

\subsection{Graphics Rendering}

Graphics rendering is critical for creating immersive VR environments, encompassing lighting, materials, textures, and visual effects. 
Both Unreal Engine and Unity employ advanced techniques to achieve high-quality visuals, with notable differences in their implementation.

\subsubsection{Lighting}

Lighting is essential for enhancing realism and depth in virtual environments. 
\textbf{Unreal Engine} leverages its \textit{Lumen} system, which delivers real-time global illumination and indirect specular reflections. 
This technology enhances the fidelity of large-scale and detailed scenes, providing a lifelike and engaging experience \cite{tan2024mastering}.

\textbf{Unity}, on the other hand, employs the \textit{High Definition Render Pipeline (HDRP)}\footnote{HDRP is a graphics pipeline tailored for high-end hardware, enabling advanced lighting and effects \cite{UnityHDRP}.}, which supports physically-based rendering and sophisticated lighting effects. 
Although achieving results comparable to Lumen often requires additional configuration, HDRP enables developers to produce high-quality lighting tailored to their project needs \cite{morse2021gaming}. 

Both engines simplify rendering complex scenes, reducing the burden on developers.

\subsubsection{Shaders}

Shaders are vital in defining object surfaces by influencing colors, textures, and lighting. \textbf{Unreal Engine} uses a physics-based material model with parameters such as \textit{BaseColor}, \textit{Metallic}, and \textit{Roughness} to accurately capture material structure and properties \cite{karis2013real}. 
It is further complemented by advanced \textit{Ray tracing} capabilities and the \textit{Nanite} system, which supports pixel-scale rendering for intricate environments \cite{pursiainen2024reaching, unreal_nanite}. 
These features make Unreal Engine a preferred choice for projects requiring high geometric detail and realism.

Unity offers flexibility through its \textit{Scriptable Render Pipelines (SRPs)} that allow developers to customize the rendering pipeline for specific project needs, including the \textit{Universal Render Pipeline (URP)} and HDRP, which are built on top of the SRP framework \cite{unityPerformanceProfilingTools}.
Shader creation is simplified using \textit{Shader Graph}, a visual tool that eliminates the need for coding. In contrast, experienced developers can use \textit{High-Level Shading Language (HLSL)} for custom implementations \cite{hasu2018fundamentals}. 
Unity’s flexibility and accessibility make it suitable for diverse development needs.

\subsection{Performance Optimization}

Performance optimization is crucial in VR applications to ensure consistent frame rates, minimize latency, and enhance user immersion. Both Unreal Engine and Unity offer robust tools for optimizing performance.

Real-time rendering generates graphics at a speed suitable for interaction, directly impacting the user’s experience. \textbf{Unreal Engine} incorporates advanced optimization tools, including \textit{Rendering Pipeline}, Optimization (RPO), which upscales from lower resolutions to reduce computational load \cite{woxler2021efficient}. The Nanite system dynamically adjusts LOD and uses Occlusion Culling to render only visible elements, significantly improving performance \cite{overton2024lods}. Additional tools, such as the Render Resource Viewer, provide detailed insights into CPU and GPU usage for fine-tuning \cite{unrealengine2025documentation5}.

\textbf{Unity} also emphasizes performance through SRPs, which allow developers to customize rendering processes for specific platforms. Techniques like Foveated Rendering reduce GPU workload by focusing high-resolution rendering on the user’s focal area \cite{petrescu2024thinking}. Unity’s Profiler and Frame Debugger provide real-time feedback on performance metrics, while the Memory Profiler helps identify bottlenecks in texture and shader usage \cite{unityPerformanceProfilingTools}. These tools enable developers to optimize applications across various hardware configurations effectively.

Both engines prioritize maintaining high frame rates (72 FPS or higher) to prevent motion sickness and ensure a smooth VR experience \cite{lavalle2023virtual}. Unreal Engine’s focus on film-quality assets and Unity’s scalability make them suitable for various VR projects, from high-end simulations to mobile applications.

\subsection{Usability}

Usability encompasses factors such as the learning curve, development efficiency, programming language support, and availability of community resources. These aspects influence the accessibility and productivity of development workflows.

\subsubsection{Learning Curve and Support}

\textbf{Unreal Engine} is renowned for its robust capabilities but has a steeper learning curve due to its complexity \cite{soni2024merits}. It supports C++ for advanced programming and Blueprints\footnote{Blueprints is a visual scripting tool in Unreal Engine that allows developers to create logic without coding \cite{unrealengine2025documentation5}.}, which caters to designers and non-programmers \cite{lee2019comparative}. However, the engine’s extensive documentation and community resources help mitigate the initial learning challenges.

\textbf{Unity} is widely regarded as more beginner-friendly, thanks to its intuitive interface and extensive online resources \cite{lee2019comparative}. It uses C\#, which integrates seamlessly with the .NET framework. Unity’s large community offers tutorials, forums, and pre-made solutions, making it an accessible platform for newcomers and professionals.

\subsubsection{Asset Ecosystem}

\textbf{Unreal Engine}’s asset ecosystem includes the Marketplace, Quixel Megascans, and Fab. These three platforms provide high-quality assets, including photogrammetry-based models and real-time rendering resources \cite{quixel2022megascans, fab_documentation}. This ecosystem enables developers to streamline workflows and reduce production costs.

\textbf{Unity}’s asset store offers a diverse library of 2D and 3D assets, scripts, and tools catering to various project needs \cite{isar2018glance}. Tools like \textit{Shader Graph} simplify shader creation, while structured documentation ensures accessibility for developers at all levels \cite{alda2023introduction}. 
% Reviewer 1 Comment 2
Quantitative benchmark data comparing both engines across multiple performance and usability metrics is presented in Table~\ref{tab:quantitative_comparison}.

\begin{table*}[h!]
\centering
\caption{Quantitative Performance Comparison between Unreal Engine and Unity}
\label{tab:quantitative_comparison}
\begin{tabular}{|p{7cm}|p{5cm}|p{5cm}|}
\hline
\textbf{Metric} & \textbf{Unreal Engine 5.3} & \textbf{Unity 2023.2 (HDRP)} \\ \hline
Average Frame Rate (Quest 2 - VR Game) & 72 FPS (with DLSS) & 72 FPS (with optimizations) \\ \hline
GPU Usage (High-fidelity Scene) & 85\% (RTX 3080) & 72\% (RTX 3080) \\ \hline
CPU Usage (Complex Scene) & 70\% (8-core CPU) & 65\% (8-core CPU) \\ \hline
VRAM Consumption (4K Scene) & 11.5 GB & 7.2 GB \\ \hline
Time to First Frame (Cold Start) & 12.3 sec & 7.8 sec \\ \hline
Average Build Time (Medium Project) & 8.4 min & 6.1 min \\ \hline
Learning Curve (Hours to Prototype) & $\sim$15--20 hrs & $\sim$8--12 hrs \\ \hline
Asset Load Time (Heavy Scene) & 5.8 sec & 4.2 sec \\ \hline
\end{tabular}
\end{table*}

%----------------

\subsection{Comparative Analysis Results Summary}\label{CA}

Overall, both engines excel in usability, with Unreal Engine providing advanced features for experienced teams and Unity offering an approachable environment for developers of all skill levels. Table \ref{tab:comparison} provides a high-level comparison of the two engines based on essential features.

\begin{table*}[h!]
\centering
\caption{Comparative Analysis of Unreal Engine and Unity}
\label{tab:comparison}
\begin{tabular}{|p{2.4cm}|p{7.3cm}|p{7.3cm}|}
\hline
\textbf{Feature} & \textbf{Unreal Engine} & \textbf{Unity} \\ \hline
\textbf{Graphics Rendering} & Advanced rendering with Lumen, Nanite, and PBR for high visual fidelity. Supports global illumination, HDR, and ray tracing. & Relies on Scriptable Render Pipeline (URP and HDRP). It is less performant for high-end rendering but offers flexibility for simpler projects. \\ \hline
\textbf{Usability} & Steeper learning curve due to C++ and unique conventions. Blueprints simplify some tasks but require advanced knowledge for optimal use. & Beginner-friendly with an intuitive interface and vast documentation. Occasional challenges with error messages and crowded UI. \\ \hline
\textbf{Asset Store} & Fab asset store integrates multiple platforms (e.g., Sketchfab, Quixel). Launched in 2024 and rapidly growing. & Extensive Unity Asset Store has a wide range of assets, including 3D/2D models, templates, AI tools, and more. \\ \hline
\textbf{Hardware Requirements} & Higher requirements for optimal performance, including RTX 2000 series or better for ray tracing and 32 GB of RAM. & More lenient requirements. It can run on lower-end devices with minimal hardware (e.g., 1 GB RAM for mobile). \\ \hline
\textbf{Cross-Platform Support} & Supports major platforms but with a focus on high-performance devices. & Extensive cross-platform support for 17 platforms, including PCs, consoles, mobile, and WebGL. \\ \hline
\textbf{Performance Optimization} & Nanite reduces computational complexity, enabling efficient rendering of high-polygon models. & Flexible optimization options for mid-range and low-end devices through URP and custom scripts. \\ \hline
\textbf{Target Audience} & Suited for projects requiring cutting-edge visuals and developers with advanced technical skills. & Ideal for beginners, small teams, and projects prioritizing accessibility and rapid development. \\ \hline
\textbf{Customization and Flexibility} & Extensive options for customization, but advanced features require significant expertise in C++ and Blueprints. & Highly flexible, with support for custom scripts using C\# and extensive third-party integrations. \\ \hline
\end{tabular}
\end{table*}

\subsubsection{Unreal Engine}

Unreal Engine stands out for its cutting-edge graphics capabilities, making it the preferred choice for VR projects demanding ultra-realistic visual fidelity. Its innovative systems, such as Lumen for global illumination and Nanite for virtualized geometry, provide developers with tools to achieve unparalleled realism and detail \cite{UnrealEngine5ReleaseNotes}. 

\textbf{Lumen and Lighting.} Unreal Engine 5's Lumen system enables real-time global illumination and dynamic lighting, enhancing immersion in VR environments. Its ability to handle infinite bounces of diffuse reflections allows developers to simulate realistic lighting in complex scenes without precomputing lightmaps \cite{tan2024mastering}. The system's software and hardware ray tracing integration also offers scalable solutions for devices with varying GPU capabilities.

\textbf{Nanite Geometry System.} Nanite's advanced virtualized geometry allows developers to use highly detailed assets without manual optimization, dynamically adjusting levels of detail (LOD) based on the viewer's distance. This feature reduces development time while maintaining graphical fidelity, making it ideal for VR applications that benefit from detailed environments \cite{overton2024lods}.

\textbf{Performance Challenges.} Unreal Engine's Physically Based Rendering (PBR) further enhances realism by simulating how light interacts with surfaces based on real-world physics. PBR incorporates base color, roughness, and metallic properties, ensuring materials appear consistent and lifelike under different lighting conditions \cite{tuliniemi2018physically}.

However, these advanced visual features come at a cost. Unreal Engine requires high-end hardware to fully leverage its features. Ray tracing and Nanite, while enhancing visual quality, can negatively impact performance on specific VR devices \cite{li2024real}. Unreal Engine demands GPUs such as NVIDIA's RTX 2000 series or higher and significant RAM resources. This poses a challenge for developers targeting devices with limited computational power \cite{UnrealEngineSpecs}.

\textbf{Development Complexity.} Unreal Engine's dual approach to development—Blueprints and C++—offers flexibility but introduces a steep learning curve. Blueprints simplify prototyping, but advanced optimizations and resource management require proficiency in C++ and familiarity with Unreal's architecture. This makes the engine more suitable for experienced developers working on high-budget or technically demanding VR projects \cite{UnrealEngine5ReleaseNotes}.

\textbf{Ecosystem and Assets.} The newly launched Fab asset store integrates resources from multiple platforms, including Sketchfab and Quixel, providing a rich repository of assets for developers. This expansion enhances Unreal's ecosystem, making it easier to find high-quality assets for VR environments \cite{epicgames2023unreal}.

\subsubsection{Unity}\label{U}

Unity's strengths lie in its accessibility and adaptability, making it a versatile tool for VR development across diverse platforms. Its emphasis on usability and extensive asset store caters to developers with varying levels of expertise, from beginners to seasoned professionals.

\textbf{Ease of Use.} Unity's intuitive interface, comprehensive documentation, and extensive tutorials make it a beginner-friendly choice. The scripting environment, primarily based on C\#, is easier to learn than Unreal Engine's C++. While occasional UI clutter and ambiguous error messages may hinder workflows, Unity's overall usability remains a significant advantage \cite{mercan2017evaluating}.

\textbf{Cross-Platform Support.} Unity excels in cross-platform compatibility, supporting 17 platforms, including mobile devices, PCs, consoles, and WebGL\footnote{WebGL is a JavaScript API that enables rendering interactive 2D and 3D graphics within web browsers without requiring plugins.}. This versatility allows developers to deploy VR applications across a broad audience, ensuring accessibility and scalability \cite{unity_multiplatform}.

\textbf{Rendering and Performance.} Unity's Scriptable Render Pipeline (SRP), which includes the Universal Render Pipeline (URP) and High Definition Render Pipeline (HDRP), provides flexibility for rendering optimization. While not as performant as Unreal's Lumen and Nanite systems for high-end visuals, SRP excels in projects targeting mid-range or mobile devices. Developers can tailor rendering pipelines to match specific hardware constraints, ensuring consistent performance across platforms \cite{soni2024merits}.

\textbf{Asset Store Ecosystem.} Unity's Asset Store offers an extensive library of assets, ranging from 3D models and audio resources to advanced tools like AI integrations and Web3\footnote{Web3 refers to the decentralized web powered by blockchain technology, enabling user-owned digital assets, smart contracts, and decentralized applications (dApps).} plugins. This robust ecosystem reduces development time by providing prebuilt components, enabling smaller teams to focus on creativity and functionality \cite{unityassetstore}.

\textbf{Hardware Requirements.} Unity's lower hardware requirements make it accessible to developers targeting devices with limited processing power. Unlike Unreal Engine, Unity can run on lower-end hardware, making it a preferred choice for mobile VR applications and projects with limited budgets. 

\begin{figure}
    \centering
    \includegraphics[width=1\linewidth]{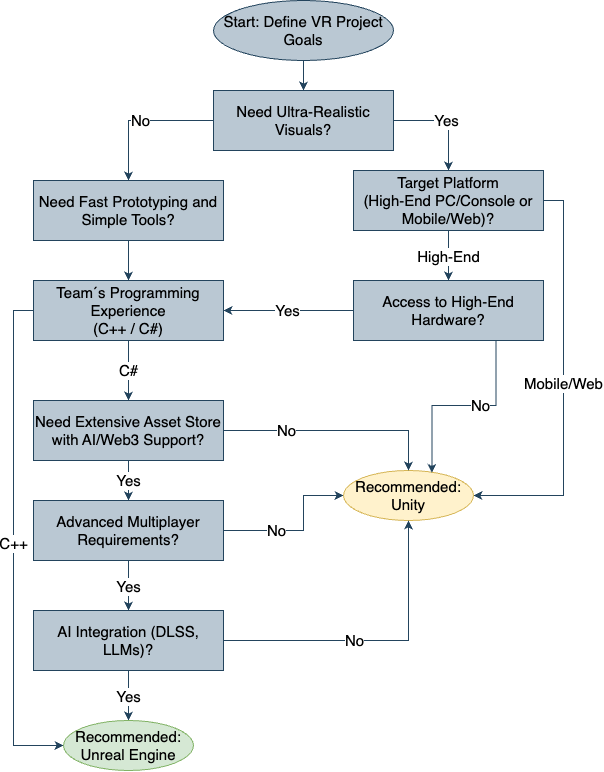}
    \caption{Flow chart of an engine selection based on the specific technical requirements as priority.}
    \label{fig:requirementsChoose}
\end{figure}

The final decision based on technical requirements is illustrated in a workflow in Figure~\ref{fig:requirementsChoose}.

\section{Case Studies in Virtual Reality Development}\label{ACS}

This section explores the application of Unreal Engine and Unity in large-scale virtual reality (VR) projects, highlighting their respective strengths, challenges, and workflows. Two exemplary case studies, the VR adaptation of \textit{Resident Evil 4} and a \textit{Virtual Museum Project}, illustrate how these engines are utilized for different objectives: immersive gaming and functional simulations.

\subsection{Resident Evil 4 VR: Unreal Engine in Action}

\textit{Resident Evil 4}, a landmark in gaming history, underwent a VR adaptation exclusively for the Quest 2 platform in 2021. The project, celebrated for its innovative gameplay and narrative, won the "Best VR/AR Game" award at The Game Awards \cite{meta_resident_2021}. This adaptation leveraged Unreal Engine’s advanced capabilities to transition a classic game into a modern VR experience while maintaining its legacy.

The adaptation faced unique challenges, including preserving the original game’s gameplay and atmosphere while introducing VR-specific features. Developers began by integrating logic code from the original game to streamline core mechanics such as enemy behavior and collision detection. Unreal Engine’s Blueprints and C++ implemented VR-specific functionalities, including intuitive aiming mechanics, controller interactions, and item handling \cite{ivey_how_2022}.

Visual fidelity was a primary focus. Textures were recreated at resolutions up to 10 times higher than the original, and lighting systems were overhauled using Unreal Engine’s Lumen and Nanite technologies. These enhancements created an immersive visual environment suitable for VR. To address performance demands, the Vulkan\footnote{Vulkan is a low-overhead, cross-platform graphics and compute API that provides high-performance access to modern GPUs.} graphics driver and custom tools were employed to optimize geometry and collision detection, achieving a stable frame rate of 72 FPS at high resolutions \cite{ivey_how_2022}.

%Reviewer 2 Comment 1 
The decision to use Unreal Engine stemmed from its robust capabilities in rendering complex scenes and handling high-resolution assets without sacrificing frame rate, a critical requirement for VR immersion in AAA titles. The integration of Lumen and Nanite allowed developers to push visual fidelity beyond what Unity could offer at the time, while its C++ and Blueprint system provided the flexibility to integrate legacy game logic with VR-specific interactions. These features made Unreal Engine the natural choice for a visually demanding, performance-sensitive project like Resident Evil 4 VR.
%---------

\begin{figure}[h!]
    \centering
    \includegraphics[height=12cm]{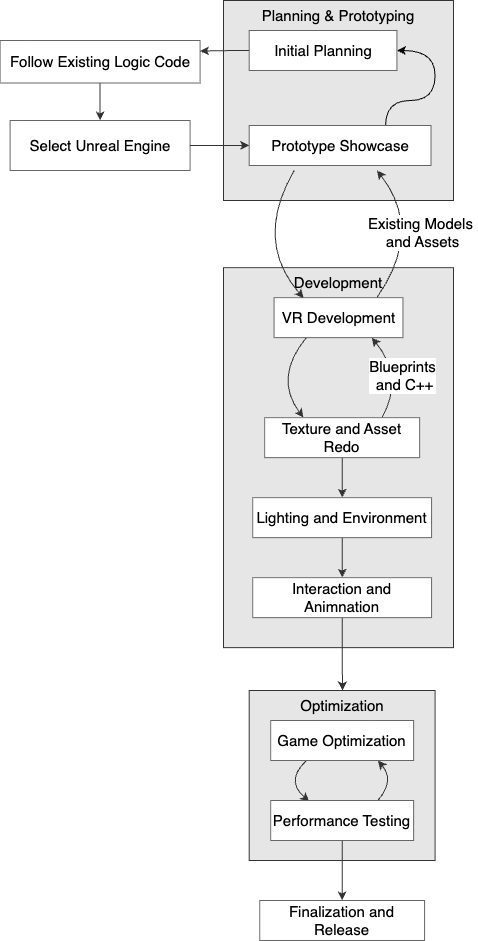}
    \caption{Workflow for the development of Resident Evil 4 VR using Unreal Engine.}
    \label{fig:work}
\end{figure}

Figure \ref{fig:work} illustrates the iterative workflow employed during the project, showcasing the adaptability of Unreal Engine to meet the rigorous demands of VR development.

\subsection{Virtual Museum Project: Unity's Role in Functional Simulations}

Unity has become a preferred choice for simulation-based VR projects due to its accessibility and flexibility. A Virtual Museum Project exemplifies Unity’s strengths in creating functional, high-precision VR applications \cite{sequeira2024design}. The project prioritized usability and realism, aiming to provide interactive educational experiences.

The development process began with selecting Unity as the primary engine for its ease of use and robust toolset. Blender was used for 3D modeling, while Unity’s \textit{XR Interaction Toolkit}\footnote{The XR Interaction Toolkit package is a high-level, component-based interaction system for creating VR and AR experiences.} streamlined the integration of interactive features. Predefined components such as the \textit{XR Ray Interactor} and \textit{XR Direct Interactor} facilitated realistic interactions, enabling users to manipulate objects and explore exhibits seamlessly.

Functional elements, including user verification systems, custom UI navigation, and device-specific adjustments, were implemented using Unity’s SRP. Interactive scaling and dynamic lighting enhanced the visualization of exhibits, creating a more engaging user experience \cite{isar2018glance}.

\begin{figure}[h!]
    \centering
    \includegraphics[height=12cm]{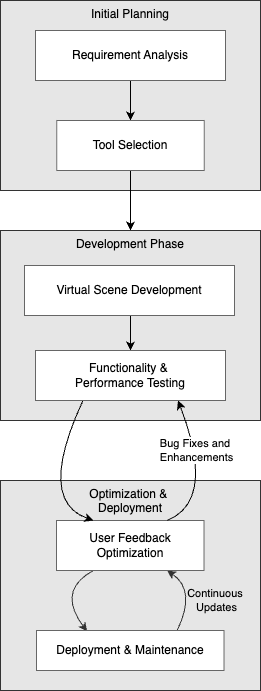}
    \caption{Workflow for the Virtual Museum Project using Unity.}
    \label{fig:museum}
\end{figure}

The project’s testing phase included multiple categories: usability, functionality, compatibility, performance, and security testing. Beta testing provided invaluable user feedback, which informed iterative improvements. Figure \ref{fig:museum} outlines the streamlined workflow for this Unity-based project.

Unity was selected due to its lightweight architecture, ease of use, and extensive XR tools, which were ideal for a project focused on educational interactivity rather than high-end visuals. The rapid prototyping capabilities provided by Unity’s XR Toolkit, combined with accessible scripting in C\# and a flexible UI framework, enabled efficient implementation of multi-platform, user-focused features. Unity’s asset store also offered quick integration of exhibit elements, further reducing development time. These factors made Unity particularly well-suited for simulation and education-oriented VR experiences.

\subsection{Insights and Implications}

The case studies of \textit{Resident Evil 4 VR} and the \textit{Virtual Museum Project} demonstrated the versatility and effectiveness of Unreal Engine and Unity in large-scale VR development. Unreal Engine’s advanced visual capabilities and optimization tools make it ideal for immersive gaming experiences, while Unity’s user-friendly environment and modular tools are well-suited for functional simulations.

These projects highlight the importance of tailored workflows, iterative testing, and leveraging engine-specific features to meet distinct project requirements. The insights gained underscore the potential of both engines to overcome technical challenges and deliver high-quality VR experiences, setting benchmarks for future development in this rapidly evolving field.

\section{Challenges and Trade-offs}\label{CA}

Developing VR applications is a complex endeavor that involves navigating various challenges and trade-offs. Game engines like Unreal Engine and Unity differ significantly in their resource requirements, licensing models, and scalability. These differences can profoundly impact the development process, influencing everything from hardware investment to the final user experience. This section delves into these critical aspects, highlighting how Unreal Engine and Unity address these challenges to meet the diverse needs of developers.

\subsection{Resource Requirements}

The hardware requirements for VR development are a significant consideration, as they directly affect the accessibility and feasibility of using a particular game engine. \textbf{Unreal Engine}, particularly in its latest version, 5.5, is known for its high hardware demands, which reflect its focus on delivering cutting-edge graphical fidelity and performance. The official recommendations for Unreal Engine 5.5 include a graphics card that supports DirectX 11 or 12 with at least 8 GB of VRAM, 32 GB of RAM, and a CPU with a clock speed of 2.5 GHz or higher. Moreover, to fully leverage advanced features like Lumen Global Illumination and Reflections, a high-performance graphics card such as the NVIDIA RTX-2000 series, AMD RX-6000 series, or newer is required \cite{unrealengine2025documentation5}. Earlier versions of Unreal Engine, such as Unreal Engine 4.2, had slightly lower requirements, needing only a GPU equivalent to the NVIDIA GeForce GTX 970 \cite{EpicGamesUE4Hardware2023}. The substantial hardware demands of Unreal Engine underscore its commitment to achieving superior visual effects. Still, they may also challenge developers with limited hardware resources, particularly indie developers or small studios.

In contrast, \textbf{Unity} is designed to be more accessible, with significantly more lenient hardware requirements. Unity 6, for example, can run on any GPU compatible with DX10, DX11, or DX12 on Windows platforms. The minimum requirements for mobile devices, such as Android phones, are as low as 1 GB of RAM. Furthermore, Unity supports platforms with similarly modest hardware needs, including game consoles and web platforms. For embedded systems running Linux, Unity requires only 1 GB of RAM, a Dual-core x64 or ARM64 CPU, and an OpenGL ES 3 or Vulkan 1.1 capable GPU \cite{unity2025requirements}. These minimal requirements allow Unity to run efficiently on mid to low-end devices, making it an ideal choice for mobile and lightweight development environments. This accessibility is one of Unity's key strengths, enabling a broader range of developers to create VR applications without needing high-end hardware.

The resource requirements of both engines are also influenced by the compatibility between their features and the specific needs of the game design. For instance, \textbf{Unreal Engine}’s Nanite technology minimizes computational complexity by simplifying the rendering of distant objects, optimizing performance while maintaining high visual quality. In contrast, \textbf{Unity} may require uniform calculations for all objects, regardless of distance, which can increase computational demands for specific scenarios. As a result, Unreal Engine may deliver similar visual effects with greater efficiency in resource-intensive projects \cite{vsmid2017comparison}. This difference in approach highlights the importance of selecting the right engine based on the project's specific requirements.

\subsection{Licensing Models}

Licensing models are another critical factor that developers must consider when choosing a game engine. \textbf{Unreal Engine} employs a tiered pricing strategy to cater to developers of varying scales and budgets. For individuals or organizations generating less than USD 1 million in gross revenue or for use in educational and academic contexts, Unreal Engine provides free access to its source code, platforms, and features, including forums and documentation. For developers exceeding USD 1 million in revenue, two licensing options are available: a Royalty-based model, which charges 5\% of gross revenue above \$1 million, or a Seat-based option, costing EUR 2,050.37 per seat annually. This flexible approach reduces entry barriers for small developers and educators while providing scalable options for larger organizations \cite{unrealengine2025license}.

\textbf{Unity}, on the other hand, offers four distinct licensing tiers to accommodate various developer needs. The Personal License is free but has limitations, such as no access to Unity's source code and restrictions on deploying games to consoles. The Unity Pro Plan, available for professional developers at €2,030.00 per year, includes game deployment permissions and additional features. The Unity Enterprise Plan provides enhanced services and technical support, with pricing available upon request. Lastly, the Unity Industry Plan, aimed at specialized and large-scale industry applications, offers an advanced toolkit for €4,554.00 per year \cite{unity_compare_plans}.

Both licensing models reflect the distinct philosophies of Unreal Engine and Unity. Unreal Engine’s royalty-based option incentivizes small developers to adopt its platform with minimal upfront costs, while its Seat-based option caters to larger studios seeking predictable expenses. Unity’s tiered plans emphasize scalability and accessibility, with a free option for hobbyists and more comprehensive plans for professional developers and enterprises. These strategies allow developers to select the model that best aligns with their financial and project-specific requirements.

\subsection{Scalability}

Scalability is crucial for VR projects, particularly those involving multiplayer functionality. Both Unreal Engine and Unity offer robust frameworks to support multi-user experiences, but their approaches and capabilities differ. 

Multiplayer functionality in VR typically involves two primary methods: local multiplayer and networked multiplayer. However, local multiplayer is generally unsuitable for VR, as VR inherently demands individualized displays and experiences \cite{EpicGamesNetworking2022}. On the other hand, networked multiplayer uses a server-client architecture to enable scalable multi-user gameplay. In this model, the server acts as the authoritative entity, ensuring game logic and state integrity. Clients send input data to the server, which processes and validates it before broadcasting updated game states back to the clients. This approach ensures synchronization, consistency, and fairness among players \cite{pennanen2022virtual, EpicGamesNetworking2022, kiselev2025advanced}.

\textbf{Unreal Engine} offers a robust server-client architecture with multiple entity types for multiplayer gaming, including standalone, client, listen server, and dedicated server \cite{sobchyshak2025}. The standalone mode is designed for local multiplayer gameplay, while the client mode only connects to the server without running any server logic. Listen servers act as both a server and a local player. In contrast, dedicated servers are optimized by excluding functions like graphics rendering, sound processing, and input handling to focus entirely on server logic. Unreal Engine utilizes an actor replication scheme to synchronize client-server states efficiently.
Additionally, the server only processes visible and relevant game data, ensuring resource optimization \cite{levchenko2020development, kiselev2025advanced}. Unreal Engine defines detailed authority and roles for multiplayer games, offering fine-grained control over player access and game logic. For developers seeking enhanced multiplayer capabilities, third-party plugins like VR Expansion Plugin provide tailored support for VR-specific multiplayer experiences \cite{VRExpansionPlugin}.

\textbf{Unity} provides various frameworks to implement multiplayer functionality, such as \textit{Photon Unity Networking (PUN)} and \textit{Mirror}, which are widely used in the Unity ecosystem. \textit{Mirror}, an open-source framework, is particularly popular due to its performance, flexibility, and support for Unity LTS versions. \textit{PUN}, Unity’s cloud-based networking framework, uses a client-server model and introduces concepts like rooms and lobbies to simplify the creation of multiplayer experiences. Players can join or create rooms for shared gameplay, with PUN handling data synchronization and player matchmaking. Other frameworks, such as \textit{MLAPI}\footnote{MLAPI (Mid-Level API) is a networking library that provides high-level abstractions for handling network communication and synchronization.} and DarkRift2\footnote{DarkRift 2 is a high-performance, multithreaded networking library for Unity, designed to provide low-latency and scalable solutions for multiplayer game development.} offers additional options for developers, varying in terms of stability, scalability, and cost, enabling developers to choose the best fit for their specific project needs \cite{aaltonen2022networking}.

\begin{figure}
    \centering
    \includegraphics[width=1\linewidth]{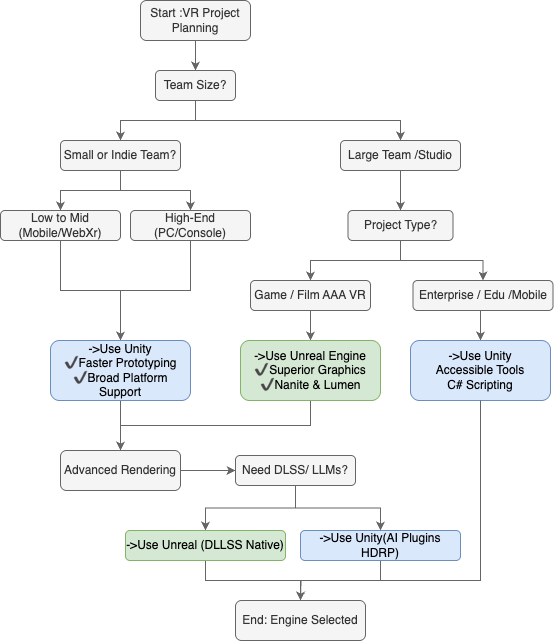}
    \caption{Flow chart selection of an Engine based on the project's scalability as priority.}
    \label{fig:goal selection}
\end{figure}

Both Unreal Engine and Unity offer powerful tools for VR development, but they differ significantly in their resource requirements, licensing models, and scalability. These differences highlight the importance of carefully considering the specific needs of a project when selecting a game engine. Whether prioritizing high-end graphical fidelity, accessibility, or multiplayer scalability, developers must weigh these factors to choose the engine that best aligns with their goals and constraints. Figure~\ref{fig:goal selection} illustrates a workflow for a proper engine selection based on the project's goal and constraints. 

\section{Future Directions}\label{F}

As Virtual Reality (VR) technology advances, emerging trends and innovations in AI, rendering techniques, and game engine usability are shaping the future of VR development. The increasing integration of AI-driven tools, such as Deep Learning Super Sampling (DLSS)\footnote{Deep Learning Super Sampling is a real-time rendering technology developed by NVIDIA that leverages AI to upscale lower-resolution images to higher resolutions, improving visual fidelity while maintaining high frame rates \cite{kapse2021overview}.} and Large language models (LLMs) are transforming traditional workflows, reducing development time, and enhancing performance. Simultaneously, ongoing improvements in game engines focus on usability, accessibility, and efficiency to support developers across various experience levels. This section explores these advancements and their potential impact on the future of VR development.

\subsection{Emerging Trends in VR Development Tools}

Integrating AI into VR development is revolutionizing game performance and content creation \cite{ozkaya2025}. DLSS exemplifies this shift by utilizing AI-based upscaling to produce high-resolution graphics while maintaining optimal performance. It reconstructs low-resolution images, analyzes frame sequences with motion data, and generates high-quality frames to enhance visual clarity. This is particularly significant for VR applications, where high frame rates are essential for a smooth and immersive experience. While Unity supports DLSS only through the HDRP, Unreal Engine provides native support, with DLSS 3 plugins fully integrated into versions 5.4, 5.3, and 5.2. The anticipated DLSS 4 update is expected to further optimize VR rendering, reducing latency and improving frame interpolation techniques \cite{nvidia_dlss}.

Beyond rendering, AI is also revolutionizing scene design and content generation. LLMs enable dynamic and context-aware environment creation, automating tasks that traditionally require significant manual effort. These advancements streamline workflows by allowing procedural content generation, real-time object placement, and intelligent world-building. Such AI-assisted design tools accelerate development and empower creators to generate complex, interactive environments with minimal programming effort. Recent studies have demonstrated collaborative and editable scene generation techniques, further opening possibilities for personalized and adaptive gaming experiences \cite{wei2024collaborative,wei2024editable}.

In parallel, real-time ray tracing continues improving VR environments' realism and physical accuracy. By enhancing lighting calculations, reflections, and global illumination, ray tracing offers superior visual fidelity compared to traditional rasterization techniques. However, its adoption in VR has been limited due to high computational costs. With hardware advancements such as NVIDIA RTX and AMD RDNA architectures, real-time ray tracing is becoming increasingly viable for VR applications, allowing developers to implement photorealistic effects without compromising performance \cite{hoffmann2011integration, schmittler2005realtime}. These innovations collectively demonstrate how AI and hardware improvements drive the next generation of VR technology.

\subsection{Improvements in Game Engines}

Despite ongoing technological advancements, game engines face significant usability and accessibility challenges, particularly for novice developers. Research indicates that beginners often struggle with complex interfaces and domain-specific programming languages, leading to longer development times and reduced efficiency \cite{mercan2017evaluating}. The steep learning curve associated with current game engines often discourages entry-level developers from engaging in VR development, limiting the diversity of content creation in the industry \cite{slootmaker2017evaluating}.

The lack of intuitive onboarding resources is a key issue contributing to these challenges. Studies have shown that fewer than 5\% of game engines achieve optimal usability based on API comprehension metrics \cite{venigalla2021comprehension}. Many new developers encounter difficulties in finding structured learning materials that align with their skill levels, forcing them to rely on fragmented community-driven tutorials. Without standardized, beginner-friendly documentation, onboarding remains a significant barrier to entry \cite{mercan2017evaluating}.

Future game engine development must emphasize user experience (UX) improvements to address these issues. Simplifying user interfaces \cite{mehmedova2025virtual}, refining documentation, and implementing AI-powered assistance tools can significantly enhance usability \cite{paduraru2024unit}. Some engines have begun integrating AI-driven coding assistants and visual scripting tools to help bridge the knowledge gap between experienced and novice developers \cite{song2025large}. Additionally, interactive tutorials and adaptive learning systems that tailor content to the user's skill level can foster a more inclusive development environment. 

By prioritizing accessibility and usability, game engines can empower a broader range of developers, from hobbyists to professionals, to contribute to VR innovation. As the industry continues to evolve, fostering an inclusive and efficient development ecosystem will be essential for sustaining the momentum of VR advancements.

\section{Conclusion}\label{C}

This study has provided an in-depth comparative analysis of Unreal Engine and Unity, two of the most widely used game engines in VR development. By examining critical factors such as graphics rendering, performance optimization, usability, resource requirements, scalability, and licensing models, this analysis offers valuable insights into how each engine caters to different development needs. Understanding these distinctions allows developers to make informed decisions when selecting the most suitable tool for their VR projects.

\textbf{Unreal Engine} excels in delivering high-fidelity visuals, leveraging state-of-the-art technologies such as the Lumen lighting system, Nanite virtualized geometry, and PBR. These features enable unparalleled graphical quality, making Unreal Engine a strong choice for developers aiming to create visually immersive and technically sophisticated VR applications. However, these advantages come at the cost of higher hardware requirements and a steeper learning curve. Unreal Engine’s complexity and resource-intensive nature may present challenges for smaller teams, independent developers, and those with limited computational resources, making it more suitable for large-scale projects and high-performance applications.

Conversely, \textbf{Unity} offers a more accessible and flexible development environment, prioritizing ease of use and broad platform compatibility. Its intuitive interface, extensive asset store, and robust cross-platform capabilities make it a preferred option for indie developers, mobile VR applications, and projects requiring rapid prototyping. Although Unity’s graphical capabilities are not as advanced as Unreal Engine’s, its lightweight architecture and lower hardware demands ensure a smoother development experience for a wider audience. Unity’s emphasis on usability and efficiency allows developers to focus on iterative design, experimentation, and faster deployment across multiple devices.

%Reviewer 2 Comment 1 
Compared to previous literature, this study provides not only a detailed technical comparison but also actionable insights drawn from real-world case studies that demonstrate how the specific needs of a VR project—such as photorealism, system complexity, scalability, or rapid development—can influence the choice between Unity and Unreal Engine. By grounding theoretical analysis in practical examples and empirical data, this paper helps bridge the gap between research and application in VR development. 
%----------

This analysis underscores the importance of aligning game engine capabilities with project-specific requirements. While \textbf{Unreal Engine} is better suited for projects demanding cutting-edge visuals and advanced customization, \textbf{Unity} offers a more accessible entry point and streamlined development for diverse applications.

Looking toward the future, emerging technologies such as AI and hardware advancements are set to further reshape the landscape of VR development. AI-powered tools, including DLSS and LLMs, are already transforming rendering workflows, automating content generation, and enhancing interactivity. Meanwhile, real-time ray tracing, hardware acceleration, and improvements in GPU performance continue to push the boundaries of visual realism and computational efficiency. As these innovations become more widely adopted, VR development will increasingly benefit from enhanced automation, higher-quality graphics, and more streamlined production pipelines.

Despite these advancements, challenges in game engine usability remain a crucial area for improvement. Enhancing onboarding processes, simplifying user interfaces, and providing beginner-friendly learning resources will be essential in fostering a more inclusive and efficient development ecosystem. By addressing these challenges, both Unreal Engine and Unity can further empower developers, from hobbyists to large studios, to create compelling VR experiences without being hindered by steep learning curves or technical limitations.

This study is a foundation for further research and development in VR technology. By leveraging the strengths of Unreal Engine and Unity while mitigating their respective limitations, developers can optimize their workflows, enhance creative possibilities, and contribute to the ongoing evolution of immersive VR experiences. As technology advances, the future of VR development holds immense potential for innovation, accessibility, and the creation of increasingly engaging virtual worlds.

\section*{Acknowledgment}

This research was financially supported by the TUM Campus Heilbronn Incentive Fund 2024 of the Technical University of Munich, TUM Campus Heilbronn. We gratefully acknowledge their support, which provided the essential resources and opportunities to conduct this study. 

\balance
\bibliographystyle{elsarticle-num}
\bibliography{VR_Engine}

\end{document}